\documentclass[10pt,conference,letterpaper]{IEEEtran}
\IEEEoverridecommandlockouts

\usepackage{cite,comment}
\usepackage{amsmath,amssymb,amsfonts}
\usepackage{algorithmic}
\usepackage{graphicx}
\usepackage{textcomp}
\usepackage{xcolor}
\usepackage{booktabs}
\usepackage{url}
\usepackage[hidelinks]{hyperref} 
\usepackage{float}
\usepackage{pifont}
\usepackage{subcaption}

\usepackage{array} 
\begin{document}

\title{Comparative Characterization of KV Cache Management Strategies for LLM Inference\\
}

\author{
\IEEEauthorblockN{
Oteo Mamo,
Olga Kogiou,
Hyunjin Yi,
Weikuan Yu
}
\IEEEauthorblockA{
\textit{Florida State University} \\
Tallahassee, FL, USA \\
\{om21d, ok22b, hy22c, wyu3\}@fsu.edu
}
}

\maketitle

\begin{abstract}
Efficient inference with Large Language Models (LLMs) increasingly relies on Key-Value (KV) caches to store previously computed key and value vectors at each layer. These caches are essential to minimize redundant computation during autoregressive token generation, lowering computational complexity from quadratic to linear. However, the growth of KV caches has posed significant system-level challenges, particularly as model sizes increase, context lengths grow, and concurrent requests compete for limited memory resources. Even though several recent frameworks for KV cache management have emerged, their comparative trade-offs in memory consumption and inference performance have not been fully understood, especially under varying request sizes and model configurations. In this work, we conduct an empirical study of three state-of-the-art KV cache management frameworks: vLLM, InfiniGen, and H2O. These frameworks employ techniques such as tensor offloading, token eviction heuristics, and speculative scheduling to balance memory usage and performance. We evaluate their performance in terms of a range of metrics such as latency, throughput, and memory usage across a spectrum of key parameters including request rates, model sizes, and sparsity levels. Our results pinpoint the conditions for each framework to perform the best, revealing the most suitable selection and configuration of KV cache strategies under memory and performance constraints.

\end{abstract}

\vspace{-1.2em}
\section{Introduction}
Large Language Models (LLMs) derive their expressive power from the Transformer architecture \cite{vaswani2017attention}, which generates text by attending over previously seen tokens using a multi-head self-attention mechanism. During inference, this mechanism incurs quadratic computational cost if past activations are repeatedly recomputed \cite{dao2023_flashattention2}. To mitigate this issue, LLMs use the Key-Value (KV) cache to store the computed tensors, i.e., keys and values, at each layer for every input or prior decoded token. Instead of recomputing these tensors at every decoding step, LLMs reuses them from KV cache to compute attention scores for newly generated tokens, reducing computational complexity and substantially accelerating inference, particularly for long sequences and autoregressive generation. However, this efficiency gain comes at the cost of increasing memory consumption from the KV cache, which grows multi-linearly with respect to sequence length, layer depth, and head width \cite{li2024survey}. As a result, KV cache memory requirements can quickly surpass that of model parameters, especially when handling batches of requests or serving long-context models \cite{ipdps25_kv_concurrency, NEURIPS2024_KVQuant}.

In response to these challenges, multiple frameworks have emerged recently, to address the growing demands of memory efficiency, resource sharing, and scalability in LLM inference. For instance, some frameworks~\cite{OSDI24_Llumnix,SOSP23_vLLM} have tried \textit{optimizing memory allocation} to reduce internal fragmentation during inference. Other techniques focus on addressing the growth of KV cache memory, applying \textit{hierarchical placement} of KV cache across GPU and CPU memory~\cite{xiao2024infllm}, or even to disk, to alleviate GPU memory pressure~\cite{ICML23_flexgen, NEURIPS2023_H2O}. In addition, attention \textit{sparsification}~\cite{NEURIPS2023_H2O, OSDI24_InfiniGen} is important to identify, prioritize and cache only important KV cache entries. Furthermore, there are techniques that alleviate the computation complexity caused by long-context lengths during inference such as 
KV cache \textit{quantization} ~\cite{NEURIPS2024_KVQuant, NEURIPS2023_ScissorHands}, compression and budgeted KV replacement~\cite{ISCA24_ALISA, tang2024quest}. While each having their benefits, these techniques create a complex repertoire of tools for system administrators and practitioners to juggle with because their comparative performance trade-offs and effectiveness to KV cache managements are not yet fully deciphered. In fact, the success of KV cache management depends on various factors, including available GPU memory, the LLM size and the request rate and scale. It is imperative for the users to be equipped with more insights on these factors in order to gauge which frameworks can provide better balance, achieving desired inference outcomes under the constraints of their system configurations.

Toward this direction, we perform a comparative characterization of three state-of-the-art frameworks: \textit{vLLM}~\cite{SOSP23_vLLM}, \textit{InfiniGen}~\cite{OSDI24_InfiniGen}, and \textit{H2O}~\cite{NEURIPS2023_H2O} to understand their trade-offs and provide empirically-grounded guidance for practitioners. Our key contributions are as follows:
\begin{itemize}
    \item We evaluate three representative KV cache frameworks across Time-to-First-Token, decode throughput, end-to-end latency, and memory consumption (GPU and CPU) under varying batch sizes and output lengths.
    \item We characterize performance bottlenecks and scalability limits across different batch sizes, context lengths, and model scales, identifying the conditions under which each paradigm excels or degrades.  
    \item We quantify the accuracy impact of KV cache sparsification across standard benchmarks and a custom retention task, identifying  sparsity configurations that can balance memory efficiency against accuracy.
\end{itemize}
While prior work has characterized individual frameworks or specific aspects of KV caching~\cite{li2024survey, SOSP23_vLLM}, this work provides a comparative evaluation and analysis across three distinct KV cache management paradigms. 

\section{Background and Motivation}
\label{background}
Now we review the current state-of-the-art strategies for the management of KV cache and how several representative frameworks have implemented these strategies.

\subsection{LLM Architecture and Inference}
Modern LLMs are built by stacking dozens to hundreds of Transformer layers, each comprising a multi-head self-attention (MHA) module followed by a position-wise feed-forward network (FFN) and wrapped in residual connections with pre-layer-norm scaling~\cite{vaswani2017attention}. However, this stack introduces significant computational and memory burdens during inference due to the sequential nature of token generation.

Transformer LLM inference consists of two distinct stages~\cite{vaswani2017attention}: in the \textbf{prefill} stage, the model processes the input prompt of length $L_{0}$ in parallel, storing the resulting key and value vectors in the KV cache for future use. In the \textbf{decode} stage, the model generates output tokens one at a time, reusing cached KV entries rather than recomputing attention over the full prompt, reducing complexity from quadratic to linear. The KV cache thus plays a pivotal role in enabling scalable and low-latency inference. However, this efficiency comes at a cost: KV cache memory grows multilinearly with sequence length, layer depth, and head width:
\[
\text{Mem}_{KV} = D \times H_{kv} \times \bigl(L_{0} + t\bigr) \times d_{h} 
\times \text{sizeof(dtype)} \times 2
\]
where $D$ is the layer count, $H_{kv}$ the number of key/value heads, $d_{h}$ the per-head dimension, $t$ the number of tokens generated so far, and the 
factor 2 accounts for both keys and values. For OPT-6.7B ($D=32$, $H_{\mathrm{kv}}=32$, $d_h=128$) at FP16/BF16, this yields $\sim$1.0~GB per sequence at 2,048 tokens, reaching $\sim$8~GB for a batch of 8. Optimizing the storage, access, and reuse of the KV cache has therefore become a central focus in modern LLM inference frameworks.

\subsection{KV Cache Management Strategies and Representative Frameworks}
\label{subsec:kv_man}

Many strategies have been proposed to cope with the growing memory consumption of KV cache, including (1) \textit{optimized memory allocation} to reduce the fragmentation of KV cache memory, (2) \textit{hierarchical placement} of KV cache that makes use of CPU memory and even storage media, (3) \textit{sparsification} techniques that identify important tokens and prioritize caching their KV entries, and (4) \textit{quantization} techniques that reduce the precision of KV entries and other tensors (e.g., from FP16 to INT8 or INT4), achieving 2-4$\times$ memory reduction at the cost of potential accuracy degradation \cite{ICML23_flexgen}. These strategies are often complementary to each other and can be combined within a single framework.

Modern architectures mitigate KV cache overhead through \textit{Grouped Query Attention} (GQA)~\cite{ainslie2023gqa}, which shares key-value heads across multiple query heads. Llama-3.1-8B~\cite{dubey2024llama}, for example, uses 32 query heads but only 8 KV heads, reducing KV cache size by $4\times$ relative to standard MHA. However, context windows have simultaneously expanded to 128K tokens, so the KV cache management techniques discussed above remain essential for efficient LLM inference.

\begin{table}[t]
\centering
\caption{Representative KV-Cache Management Frameworks}
\label{tab:derivative_frameworks}
\begin{tabular}{>{\raggedright\arraybackslash}p{4.0cm}>{\raggedright\arraybackslash}p{4.2cm}}
\toprule
\textbf{Representative Frameworks} & \textbf{Shared Characteristics} \\
\midrule
\textbf{vLLM}, ORCA, vTensor & Paged memory, continuous batching, no sparsification \\
\midrule
\textbf{H2O}, Scissorhands, StreamingLLM, ALISA, SnapKV & Permanent eviction, attention-score selection, static or sliding-window \\
\midrule
\textbf{InfiniGen}, InfLLM, Quest, PQCache & Hierarchical storage, decode-time selection, recoverable eviction \\
\bottomrule
\end{tabular}
\vspace{-1.2em}
\end{table}

\subsection{Diverse Paradigms of KV Cache Frameworks.} 

Existing KV cache management frameworks can be broadly categorized into three paradigms based on their primary optimization strategy and eviction policy. Table~\ref{tab:derivative_frameworks} summarizes representative frameworks within each category:

\begin{enumerate}
    \item \textit{Memory Management} frameworks optimize allocation and scheduling without modifying attention computation or evicting cache entries. These frameworks retain the complete KV cache on GPU, focusing on reducing fragmentation and improving batching efficiency.
    
    \item \textit{Static Sparsification} frameworks reduce cache size through permanent token eviction based on attention scores computed during prefill. Once evicted, tokens cannot be recovered, trading potential accuracy loss for aggressive memory reduction.
    
    \item \textit{Dynamic Selection} frameworks retain the full KV cache in CPU memory and selectively transfer entries to GPU at each decode step. This approach enables recovery of previously offloaded tokens, preserving accuracy while reducing GPU memory pressure.
\end{enumerate}

We select one representative framework from each paradigm for our comparative study: vLLM (memory management), H2O (static sparsification), and InfiniGen (dynamic selection). These categories reflect primary optimization strategies rather than strict boundaries. Each of these frameworks has significantly influenced subsequent work within its paradigm and provides a well-documented, reproducible implementation. We elaborate on each below.

\textbf{vLLM.} Kwon et al.~\cite{SOSP23_vLLM} developed vLLM, a framework that uses PagedAttention, an algorithm that partitions the KV cache into fixed-size blocks, allowing them to be stored in non-contiguous GPU memory and significantly reducing fragmentation. This management of KV cache blocks allows vLLM to batch more sequences, as common prefix tokens are not duplicated but stored once and shared across requests via copy-on-write semantics. vLLM also integrates FlashAttention-2~\cite{dao2023_flashattention2} to address inefficiencies in attention computation by fusing small blocks of attention computation into a single GPU kernel for efficient data accesses within GPU SRAM, reducing traffic to GPU global memory and improving GPU utilization. While vLLM supports various KV cache optimization add-ons, we maintain default configuration, retaining the complete KV cache, as our baseline for full-accuracy inference.

\textbf{H2O.} Zhang et al. proposed Heavy Hitter Oracle (H2O)~\cite{NEURIPS2023_H2O}, a framework that exploits the observation that attention weights are highly skewed toward a small subset of tokens. H2O identifies these ``Heavy Hitters'', tokens that accumulate the highest attention scores, and retain only their KV vectors. It also forms a sliding window of recent tokens, permanently evicting all others. This approach frames KV cache eviction as a dynamic submodular maximization problem, providing theoretical guarantees on approximation quality. The selection occurs during prefill and remains static throughout decoding. This aggressive sparsification reduces the size of KV cache while maintaining reasonable accuracy.

\textbf{InfiniGen.} Lee et al.~\cite{OSDI24_InfiniGen} proposed InfiniGen, a framework that takes a fundamentally different approach by offloading the full KV cache from GPU to CPU memory while dynamically selecting which entries to transfer to GPU at each decode step. InfiniGen employs an offline \textit{skewing} phase, Singular Value Decomposition (SVD) to the query and key weight matrices, identifying dimensions that dominate attention computations. At runtime, InfiniGen maintains compact \textit{partial key} matrices on GPU that capture these dominant dimensions. At each decode step, the framework uses these partial matrices to \textit{speculate} on which tokens will receive high attention scores in the next layer, then selectively prefetches only those KV entries from CPU to GPU. This approach enables InfiniGen to adapt to changing attention patterns throughout generation.

\subsection{Motivation}

While substantial progresses have been made in developing KV cache management techniques, current understandings are typically gained from the assessment of individual frameworks or by comparing techniques within a single paradigm. A significant gap remains in understanding the trade-offs in KV cache management across different paradigms, e.g., memory management without sparsification, static eviction based on prefill-time importance, and dynamic selection with recoverable entries. Our work addresses this gap by providing a \textit{comparative and empirical} evaluation across the different approaches. By systematically varying model size, batch size, context length, and sparsity budget, we pinpoint the conditions under which each paradigm performs the best, offering practical guidance for framework selection under various memory and performance constraints.

\section{Experimental Setup}  
\label{sec:system}
\subsection{Hardware environment} 
All experiments were run on a compute node equipped with 4$\times$ NVIDIA H100 GPUs (80GB HBM3 each) and dual Intel Sapphire Rapids processors (56 cores per socket, 2 sockets per node), interconnected via NVLink and PCIe 5.0.

\subsection{Models and Datasets} 
We evaluate three models spanning different parameter scales: Llama-3.1-8B, Llama-3.1-70B \cite{dubey2024llama}, and GPT-OSS-20B \cite{openai2023gpt4}. All models employ GQA, reflecting the shift in both open-source and academic communities away from MHA toward more memory-efficient attention architectures.
 For Time-To-First-Token (TTFT) and accuracy evaluation, we use the LMSYS-Chat-1M\cite{zheng2023lmsys} dataset, which provides realistic conversational workloads with diverse prompt lengths. For stress testing under extreme sequence lengths, we stream the input from the English Wikipedia dataset (20220301.en split) \cite{wikipedia20220301en}. For accuracy evaluation, we employ six reasoning benchmarks via the \texttt{lm\_eval} harness: PIQA \cite{{bisk2020piqa}} and HellaSwag \cite{zellers2019hellaswag} (physical and commonsense reasoning), COPA \cite{roemmele2011copa} (causal inference), WinoGrande \cite{sakaguchi2020winogrande} (pronoun resolution), OpenBookQA \cite{OpenBookQA2018} and BoolQ \cite{clark2019boolq} (knowledge-intensive question answering).

\subsection{Framework Configurations and Instrumentation} For vLLM, we use version 0.15.0 with FlashAttention-2 and continuous batching enabled. vLLM's inference engine automatically applies chunked prefill for long prompts. For InfiniGen, we configure a KV budget of 0.3 and FlashAttention-2 integration is evaluated separately where noted. For H2O, we have implemented the algorithm on top of FlexGen following the original methodology, which enables sparsification across batched decoding. We configure a Heavy-Hitter ratio (HH) of 0.3. For InfiniGen and H2O, the prefill stage requires full attention score matrix computation, which exhausts GPU memory when processing multiple prompts concurrently. Our implementation therefore performs sequential prefill: prompts are processed one at a time during the prefill phase, with batched decoding enabled after all prompts have been prefilled. The configuration choices for H2O and InfiniGen are made because both achieve comparable accuracy on the benchmark suite, maintaining performance within 10\% of baseline vLLM for most datasets. This calibration ensures a fair comparison of system-level metrics under equivalent accuracy constraints. Detailed accuracy results are presented in Section~\ref{sec:sparsification}.

\section{Overall Performance Analysis}
\label{sec:overall}

We evaluate the performance of vLLM, InfiniGen, and H2O in managing the KV cache for LLM inference across three experimental dimensions: Time-to-First-Token (TTFT) for long-context prefill behavior, Batch Scaling and Resource Utilization for concurrent workload performance, and Decode Scaling with Output Length for sustained generation efficiency.

\subsection{Time-to-First-Token (TTFT) Analysis}
\label{subsec:ttft}

TTFT measures the latency from request submission to the generation of the first output token. At long context lengths, TTFT is dominated by the prefill phase, where the model processes the entire input prompt and populates the KV cache. This metric is critical for user-perceived responsiveness in interactive applications.

\subsubsection{Baseline TTFT Comparison}
\begin{figure}[t]
  \centering
  \begin{subfigure}[t]{0.49\linewidth}
    \centering
    \includegraphics[width=\linewidth]{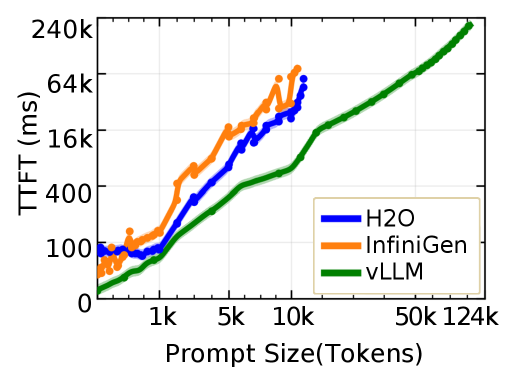}
    \caption{\centering TTFT (Llama-3.1-8B, 1xH100)}
    \label{fig:ttft_a}
  \end{subfigure}\hfill
  \begin{subfigure}[t]{0.49\linewidth}
    \centering
    \includegraphics[width=\linewidth]{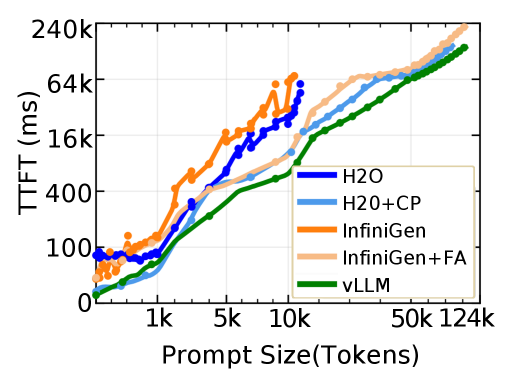}
    \caption{\centering TTFT Extended (Llama-3.1-8B, 1xH100)}
    \label{fig:ttft_b}
  \end{subfigure}

  \begin{subfigure}[t]{0.48\linewidth}
    \centering
    \includegraphics[width=\linewidth]{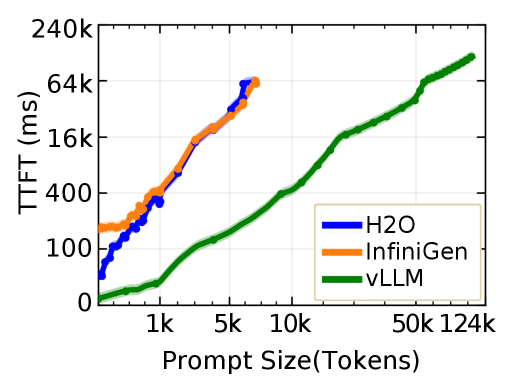}
    \caption{\centering TTFT (GPT-OSS-20B, 2$\times$H100)}
    \label{fig:ttft_c}
  \end{subfigure}\hfill
  \begin{subfigure}[t]{0.50\linewidth}
    \centering
    \includegraphics[width=\linewidth]{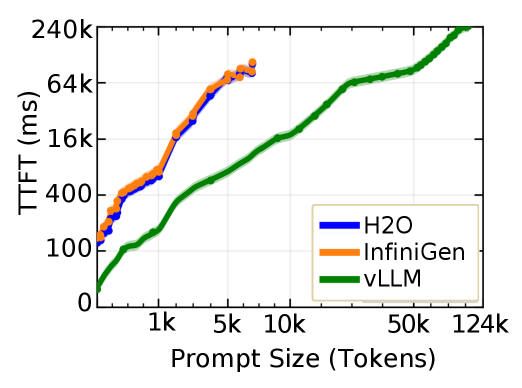}
    \caption{\centering TTFT (Llama-3.1-70B, 4$\times$H100)}
    \label{fig:ttft_d}
  \end{subfigure}
  \caption{TTFT scaling with prompt length. (a) Baseline comparison showing OOM boundaries for H2O and InfiniGen. (b) Extended TTFT with FA and CP optimizations. (c,d) Multi-GPU scaling on larger models.}
  \label{fig:ttft}
\vspace{-1.2em}
\end{figure}

vLLM maintains the lowest TTFT across all prompt lengths, successfully scaling to the full 128K context window (Figure~\ref{fig:ttft_a}). In contrast, baseline H2O and InfiniGen encounter out-of-memory failures at approximately 10-12K tokens, preventing them from processing even 10\% of the supported context length. These failures occur during prefill attention computation, where memory grows quadratically with prompt length. Thus, despite reducing KV cache memory, both frameworks remain limited by the memory cost of full-context attention during prefill.

These patterns hold across model scales (Figures~\ref{fig:ttft_c},~\ref{fig:ttft_d}), with a counterintuitive finding: OOM failures occur earlier on multi-GPU configurations despite greater aggregate memory. Under tensor parallelism, model parameters are distributed across GPUs, but the attention score computation remains local to each device and therefore does not benefit proportionally from the additional aggregate memory. Consequently, the $O(n^2)$ attention matrix can still exhaust per-device memory, causing OOM at relatively short context lengths.

\subsubsection{Extending Context Range: FlashAttention-2 vs.\ Chunked Prefill}

Two complementary strategies address prefill memory bottlenecks: FlashAttention-2 (FA) and Chunked Prefill (CP). FA fuses attention operations into a single GPU kernel that computes attention incrementally within SRAM tiles, avoiding materialization of the $O(n^2)$ attention matrix and reducing memory complexity to $O(n)$. Chunked Prefill partitions the input prompt into sequential segments of size $c$, reducing peak memory from $O(n^2)$ to $O(c^2)$ per chunk at the cost of additional kernel launch overhead and reduced parallelism. vLLM natively integrates FA and CP automatically, which contributes to its superior long-context performance as our evaluation will show. For InfiniGen and H2O, which rely on standard attention in their baseline implementations, we evaluate both FA and CP as mitigation strategies.

\begin{figure}[t]
  \centering
    \begin{subfigure}[t]{0.490\linewidth}
      \centering
      \includegraphics[width=\linewidth]{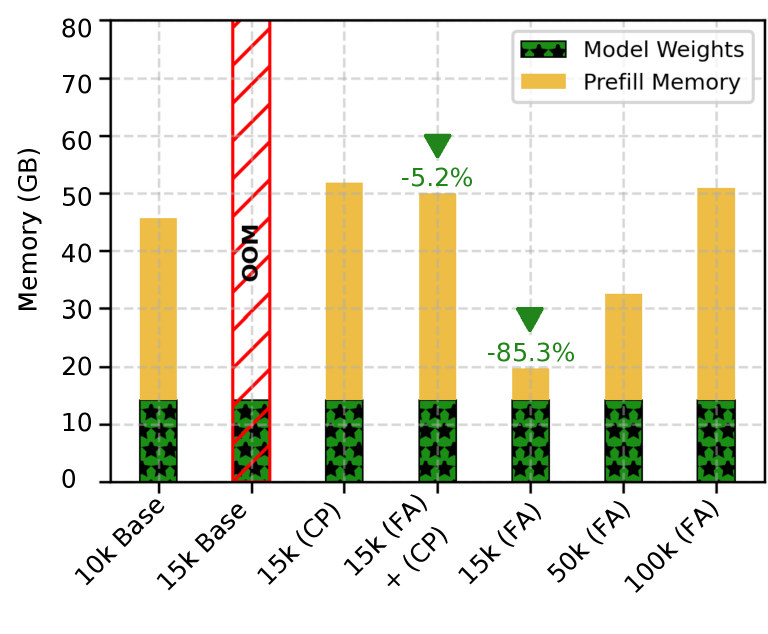}
      \caption{GPU Memory Breakdown}
      \label{fig:infinigen_FA1}
    \end{subfigure}\hfill
    \begin{subfigure}[t]{0.465\linewidth}
      \centering
      \includegraphics[width=\linewidth]{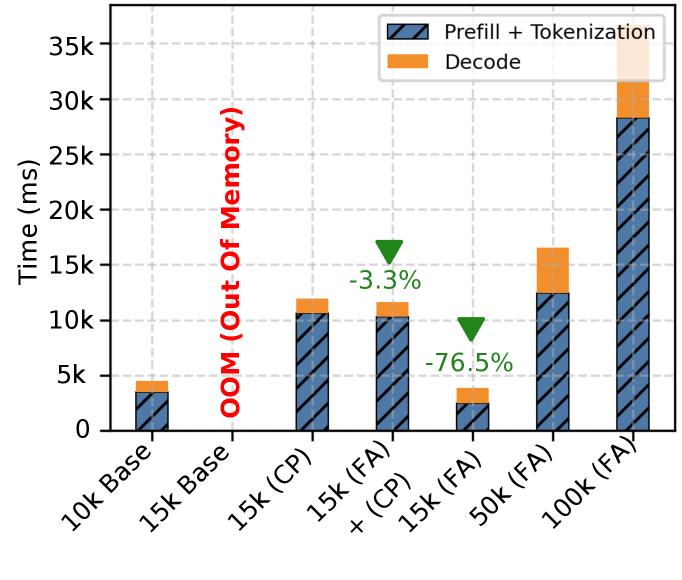}
      \caption{Latency Breakdown}
      \label{fig:infinigen_FA2}
    \end{subfigure}
  \caption{InfiniGen memory and latency breakdown across optimization strategies 
(Llama-3.1-8B). (a) GPU memory at 10K--100K token contexts. 
(b) Prefill and decode latency at equivalent context lengths. Hatched bars 
indicate OOM. Percentages show reduction relative to the comparison baseline.}
  \label{fig:infinigen_FA}
\vspace{-1.2em}
\end{figure}

\paragraph{InfiniGen} 
FlashAttention-2 integration transforms InfiniGen's long-context viability. At 15K tokens, where baseline InfiniGen fails, FlashAttention-2 reduces prefill memory by 85\% (Figure~\ref{fig:infinigen_FA1}) and latency by 77\% compared to chunked prefill (Figure~\ref{fig:infinigen_FA2}). More importantly, FlashAttention-2 enables InfiniGen to scale to the full 128K context window with near-linear latency growth, matching vLLM's operational 
range albeit at lower throughput.

\paragraph{H2O} Unlike InfiniGen, H2O does not benefit from FlashAttention-2 integration. While FlashAttention-2 can be used for the attention computation itself, H2O's Heavy-Hitter selection mechanism independently requires materializing the full attention score matrix $\text{softmax}(QK^T)$ to accumulate token importance statistics for eviction decisions. This negates FlashAttention's memory savings, as the $O(n^2)$ matrix must still be computed and stored regardless of how the attention output is calculated. H2O can instead leverage Chunked Prefill with a chunk size of up to  6k tokens, large enough to amortize chunking overhead, yet small enough to avoid OOM during per-chunk attention computation. This configuration successfully scales to the full context window (Figure~\ref{fig:ttft_b}). However, this approach introduces a fundamental algorithmic limitation: H2O's Heavy-Hitter selection relies on computing attention across the entire prompt to identify important tokens. With Chunked Prefill, tokens in different chunks do not attend to each other during prefill, meaning H2O's eviction decisions are based on incomplete attention information. This disconnect between chunks can result in the eviction of potentially important tokens that would have been identified as Heavy Hitters had full cross-prompt attention been computed. One potential approach to addressing this limitation would be to accumulate attention statistics across chunks during prefill, though the specific implementation strategy would depend on the deployment requirements and performance trade-offs acceptable for the particular workload.

\subsection{Resource Utilization and Batch Scaling}
\label{subsec:batch_scaling}

We evaluate resource utilization and performance across batch sizes from 16 to 96 on Llama-3.1-8B (300 Wikipedia-derived prompts, lengths uniformly distributed 1K-10K tokens), revealing fundamental trade-offs in how each paradigm balances memory consumption against throughput.

\begin{figure}[t]
  \centering
  \begin{subfigure}[t]{\linewidth}
    \centering
    \includegraphics[width=\linewidth]{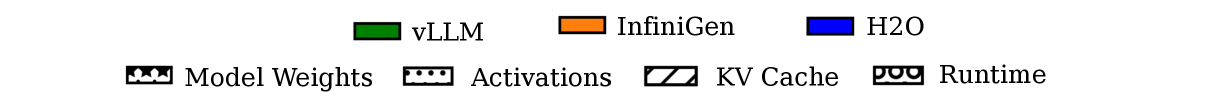}
    \label{fig:legend}
  \end{subfigure}
  \begin{minipage}{\linewidth}
    \vspace*{-1.2em}
    \centering
    \begin{subfigure}[t]{0.45\linewidth}
      \centering
      \includegraphics[width=\linewidth]{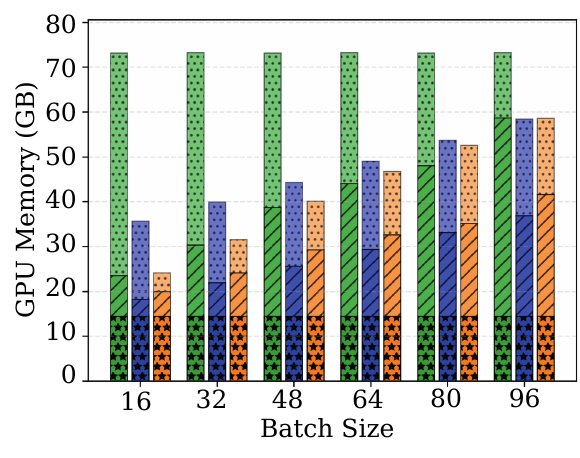}
      \caption{GPU Memory Usage}
      \label{fig:GPU_memory}
    \end{subfigure}\hspace{0.02\linewidth}
    \begin{subfigure}[t]{0.45\linewidth}
      \centering
      \includegraphics[width=\linewidth]{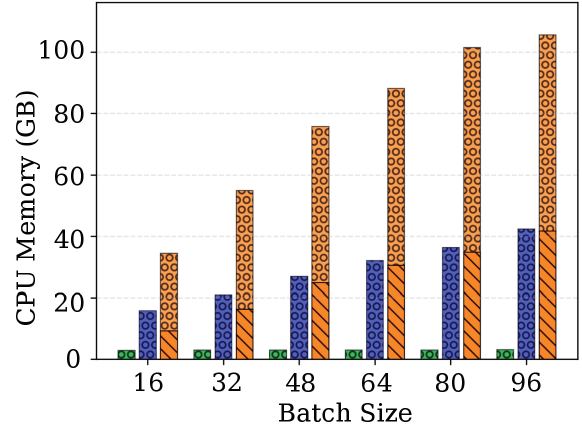}
      \caption{CPU Memory Usage}
      \label{fig:CPU_Memory}
    \end{subfigure}
  \end{minipage}
  \caption{GPU and CPU memory breakdown by component across batch sizes (Llama-3.1-8B, prompt lengths 1 to 10K tokens). Memory is decomposed into model weights, activations, KV cache, and runtime overhead.}
  \label{fig:memory_usage}
  \vspace{-1.2em}
\end{figure}
\subsubsection{Memory Utilization}

GPU and CPU memory utilization reveal distinct resource trade-offs across paradigms (Figure~\ref{fig:memory_usage}). vLLM pre-allocates GPU memory for the full context window, maintaining a constant $\sim$72GB footprint regardless of batch size, a strategy that sacrifices memory flexibility for computational efficiency. H2O takes the opposite approach: aggressive sparsification yields the lowest GPU footprint (under 60GB even at batch 96), scaling at roughly 2$\times$ lower memory than vLLM across all batch sizes. InfiniGen's hierarchical KV placement shifts the memory burden from GPU to CPU. While its GPU memory consumption is similar to H2O, CPU memory grows substantially with batch size, exceeding 100GB at batch 96, more than 2.5$\times$ the CPU memory required by either alternative. This overhead stems not only from the offloaded KV cache itself, but also from pinned memory buffers required for asynchronous CPU-GPU transfers and memory pools that facilitate rapid data movement. H2O's CPU usage primarily comes from temporarily holding evicted KV cache entries.

\subsubsection{Throughput Analysis}

\begin{figure}[t]
  \centering
    \begin{subfigure}[t]{0.450\linewidth}
      \centering
      \includegraphics[width=\linewidth]{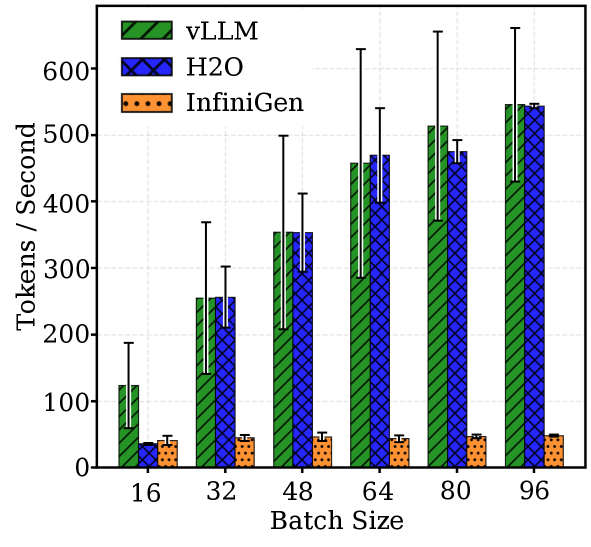}
      \caption{Decode Throughput}
      \label{fig:throughput_plot}
    \end{subfigure}\hspace{0.02\linewidth}
    \begin{subfigure}[t]{0.450\linewidth}
      \centering
      \includegraphics[width=\linewidth]{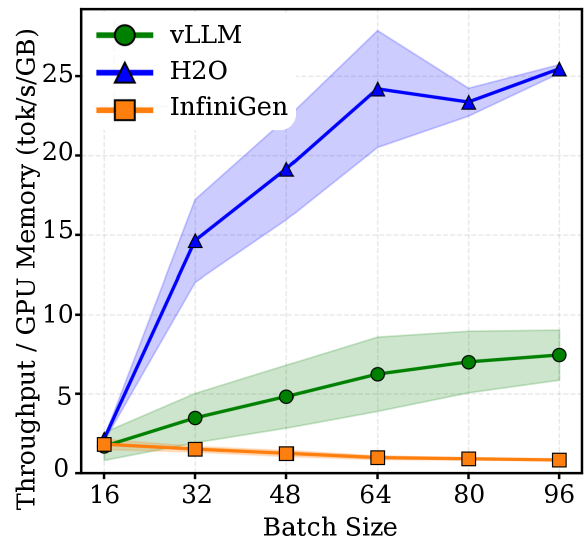}
      \caption{Memory Efficiency}
      \label{fig:throughput_gpu_plot}
    \end{subfigure}
  \caption{Decode throughput (a) and memory efficiency (b) across batch sizes. Efficiency computed as throughput divided by GPU memory consumption. Error bars and shaded regions indicate variance across runs.}

  \label{fig:throughput_memory}
   \vspace{-1.2em}
\end{figure}

Decode throughput varies significantly across paradigms (Figure~\ref{fig:throughput_plot}). vLLM and H2O both scale linearly with batch size, reaching similar peak throughput. InfiniGen, however, exhibits an order-of-magnitude lower throughput consistently across all batch sizes. This throughput gap of InfiniGen stems from per-decode-step CPU-to-GPU KV transfers, which serialize batch computation: increasing batch size adds more transfers rather than amortizing fixed overhead. When normalized by GPU memory consumption (Figure~\ref{fig:throughput_gpu_plot}), H2O emerges as the most memory-efficient option, achieving approximately 3$\times$ higher throughput per GB than vLLM. After factoring the memory consumption, InfiniGen shows poor and decreasing normalized throughput despite its reduced GPU memory footprint. 

\begin{figure}[t]
  \centering
    \begin{subfigure}[t]{0.45\linewidth}
      \centering
      \includegraphics[width=\linewidth]{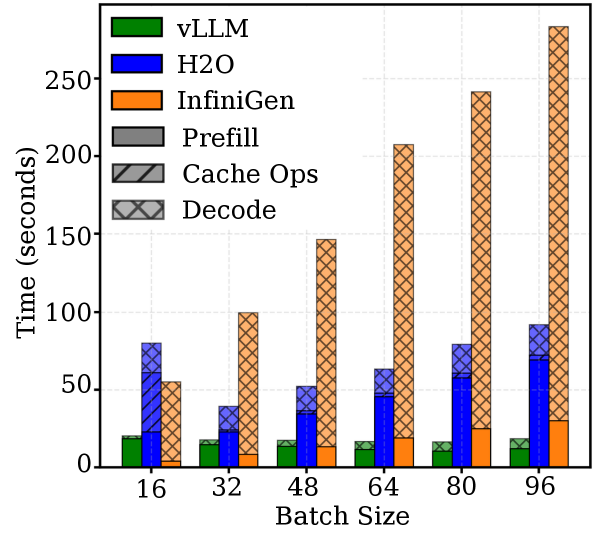}
      \caption{Walltime Breakdown}
      \label{fig:walltime_plot}
    \end{subfigure}\hspace{0.02\linewidth}
    \begin{subfigure}[t]{0.450\linewidth}
      \centering
      \includegraphics[width=\linewidth]{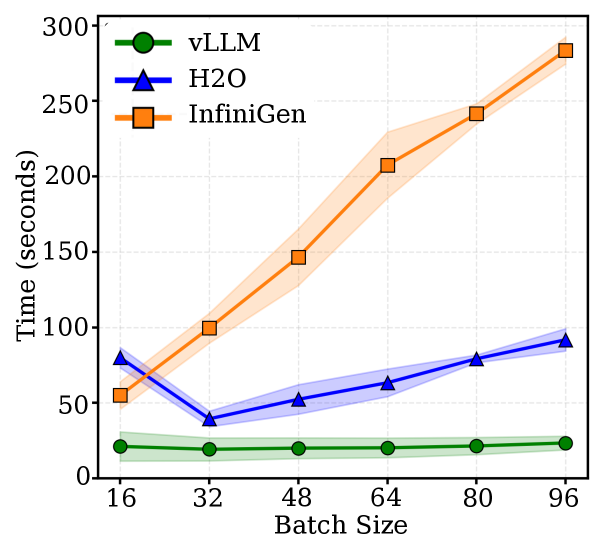}
      \caption{Total Latency}
      \label{fig:latency_plot}
    \end{subfigure}
  \caption{End-to-end latency analysis across batch sizes. (a) Walltime decomposed into prefill, cache operations, and decode phases. (b) Total latency scaling.}
  \label{fig:latency_walltime}
   \vspace{-1.2em}
\end{figure}

\subsubsection{Latency Breakdown}

Latency scaling results with increasing batch sizes further differentiate the three frameworks (Figure~\ref{fig:latency_plot}). vLLM maintains near-constant latency across all batch sizes, approximately 20 seconds. H2O exhibits moderate growth, which is due to the increasing KV cache management overhead for larger batches.
InfiniGen's latency, however, grows superlinearly, reaching 14$\times$ higher than vLLM at batch 96. The walltime breakdown in Figure~\ref{fig:walltime_plot} confirms that the time for decode-phase transfers dominates, while the prefill time remains stable.

\subsection{Decode Scaling with Output Length}
\label{subsec:decode_scaling}

While Section~\ref{subsec:batch_scaling} examined batch scaling with fixed output length, many applications require generating extended outputs, long-form articles, detailed code implementations, or multi-step reasoning chains. This section characterizes how throughput and latency evolve as output length increases from 512 to 8,192 tokens, using a fixed 10K-token input prompt.

\subsubsection{Throughput Sustainability}
\begin{figure}[t]
  \centering
  \begin{subfigure}[t]{0.45\linewidth}
    \centering
    \includegraphics[width=\linewidth]{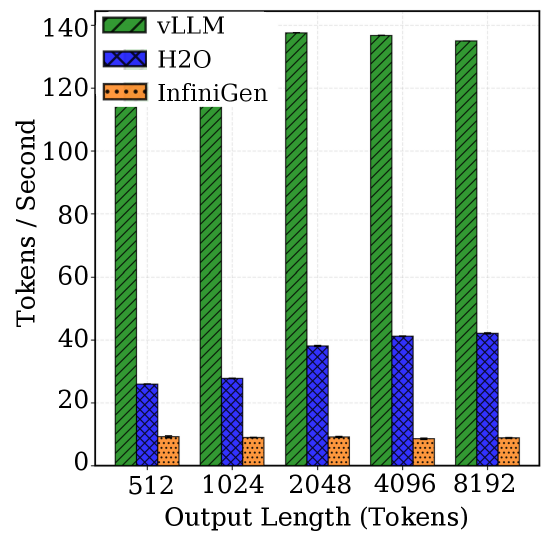}
    \caption{Decode Throughput}
    \label{fig:decode_throughput}
  \end{subfigure}\hspace{0.02\linewidth}
  \begin{subfigure}[t]{0.45\linewidth}
    \centering
    \includegraphics[width=\linewidth]{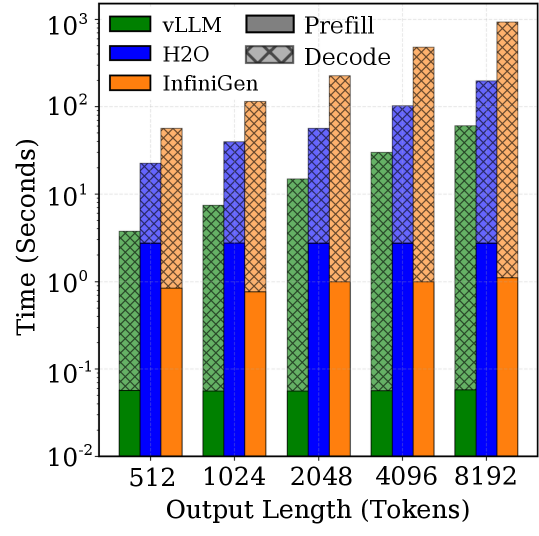}
    \caption{Walltime Breakdown}
    \label{fig:decode_walltime}
  \end{subfigure}
  \caption{Decode scaling with output length on Llama-3.1-8B. (a) Decode throughput. (b) Walltime breakdown; logarithmic scale for prefill and decode components.} 
  \label{fig:decode_scaling}
   \vspace{-1.2em}
\end{figure}

The throughput results differ significantly between frameworks across different output lengths (Figure~\ref{fig:decode_throughput}). vLLM maintains approximately 15$\times$ higher throughput than InfiniGen, which remains consistent even with longer output tokens being generated. This persistence confirms that InfiniGen's bottleneck is per-token transfer overhead rather than the startup cost that can be amortized by longer output tokens.

Both vLLM and H2O exhibit slight throughput improvements at longer outputs as fixed overheads amortize across more tokens. H2O's bounded cache size additionally prevents attention computation costs from growing with sequence length, partially offsetting its cache management overhead. InfiniGen's throughput, however, remains flat at approximately 8 tokens/second: each decode step requires speculative attention computation and selective KV prefetching, creating per-token overhead that dominates 
regardless of generation length.

\subsubsection{Walltime Breakdown}

End-to-end latency differences compound dramatically over long generations (Figure~\ref{fig:decode_walltime}). Prefill time remains constant across output lengths as expected given the fixed input prompt. Decode time, however, scales linearly with output length at vastly different rates across frameworks. At 8,192 output tokens, InfiniGen requires over 16 minutes compared to vLLM's one minute: a 17$\times$ slowdown. H2O occupies the middle ground at approximately 3 minutes, its GPU-resident computation avoiding InfiniGen's transfer overhead while cache management adds modest latency compared to vLLM. This disparity underscores that serialized CPU-GPU transfers fundamentally limit InfiniGen's decode performance regardless of generation length.

Decode scaling confirms that the performance gaps observed in batch scaling compound over extended generations. GPU-local KV cache operations remain the primary determinant of throughput sustainability, making InfiniGen's hierarchical approach viable only when GPU memory constraints are severe and latency requirements are relaxed.

\section{Accuracy Impact of Attention Sparsification}
\label{sec:sparsification}

\subsection{Accuracy Evaluation}
\begin{figure}[t]
  \centering
  \includegraphics[width=\linewidth]{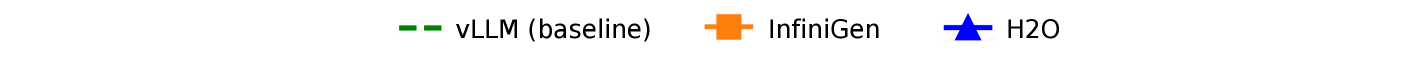}
    \vspace*{-1.2em}
  
  \begin{subfigure}[t]{0.31\linewidth}
    \centering
    \includegraphics[width=\linewidth]{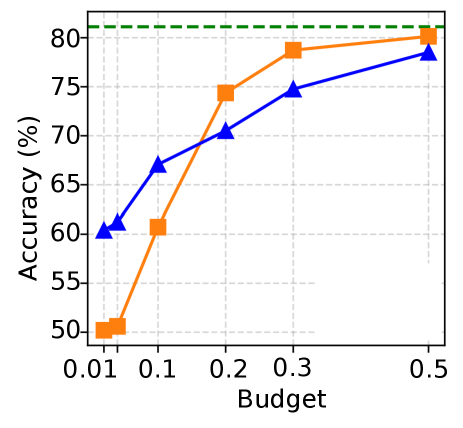}
    \caption{\centering PIQA}
    \label{fig:acc_piqa}
  \end{subfigure}\hfill
  \begin{subfigure}[t]{0.29\linewidth}
    \centering
    \includegraphics[width=\linewidth]{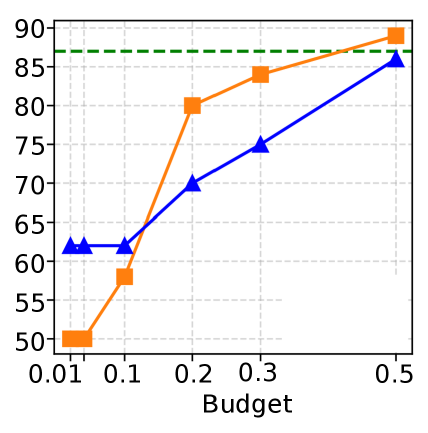}
    \caption{\centering COPA}
    \label{fig:acc_copa}
  \end{subfigure}\hfill
  \begin{subfigure}[t]{0.29\linewidth}
    \centering
    \includegraphics[width=\linewidth]{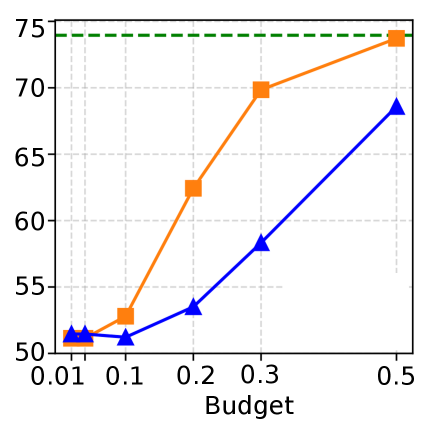}
    \caption{\centering WinoGrande}
    \label{fig:acc_winogrande}
  \end{subfigure}
  \vspace{0.6em}
  
    \begin{subfigure}[t]{0.32\linewidth}
    \centering
    \includegraphics[width=\linewidth]{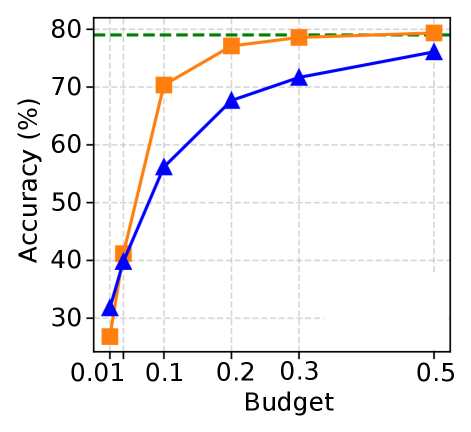}
    \caption{\centering HellaSwag}
    \label{fig:acc_hellaswag}
  \end{subfigure}\hfill
  \begin{subfigure}[t]{0.30\linewidth}
    \centering
    \includegraphics[width=\linewidth]{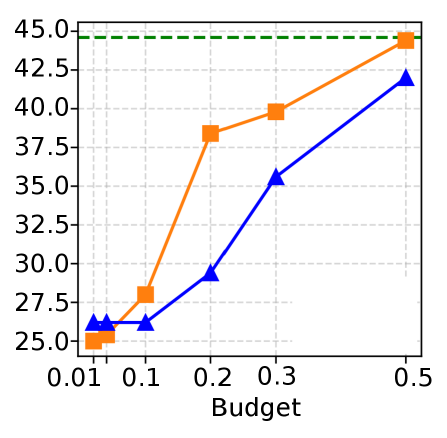}
    \caption{\centering OpenBookQA}
    \label{fig:acc_openbookqa}
  \end{subfigure}\hfill
  \begin{subfigure}[t]{0.30\linewidth}
    \centering
    \includegraphics[width=\linewidth]{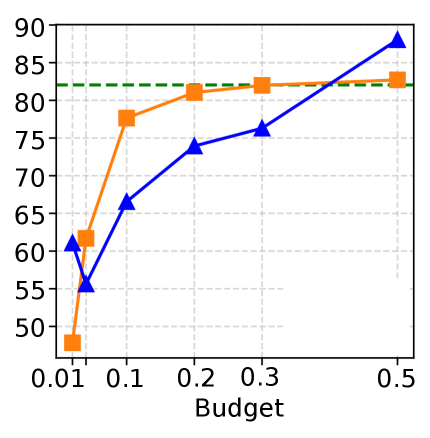}
    \caption{\centering BoolQ}
    \label{fig:acc_boolq}
  \end{subfigure}
  \caption{Accuracy comparison of InfiniGen, H2O, and vLLM}
  \label{fig:accuracy_grid}
 \vspace{-1.2em}
\end{figure}

The previous sections demonstrate that sparsification frameworks achieve substantial memory and throughput benefits compared to full-cache approaches. However, these gains are only meaningful if accuracy remains acceptable. This section evaluates the accuracy implications of H2O's permanent eviction and InfiniGen's dynamic selection strategies across standard benchmarks and a custom retention task, using vLLM as the full-accuracy baseline.
Using Llama-3.1-8B, we evaluate six log-likelihood ranking benchmarks: PIQA, HellaSwag, WinoGrande, COPA, OpenBookQA, and BoolQ. The budget parameter specifies the fraction of KV pairs each layer may retain; for example, a budget of 0.1 permits only one-tenth of the cache entries to reside in GPU memory. Figure~\ref{fig:accuracy_grid} summarizes accuracy across all six tasks.
At moderate sparsity (budget=0.5), both methods recover near-baseline accuracy. Reducing the budget to 0.1 reveals stronger and highly task-dependent degradation. InfiniGen loses approximately 6 percentage points (pp) on average, but suffers substantially larger drops on several tasks. H2O also shows pronounced sensitivity, with HellaSwag dropping 23 pp and BoolQ dropping 16 pp. At extreme sparsity (2-4\% budget), both methods approach random-guess performance on several semantic reasoning tasks. Overall, the results show that accuracy degradation under sparsification depends strongly on both the compression level and the downstream task; moderate sparsity can preserve near-baseline accuracy on some benchmarks, while aggressive sparsity produces substantial losses for both methods.

\subsection{Retention Ability}

Standard benchmarks evaluate general reasoning, but sparsification may disproportionately affect retention of specific early-context information. We design a retention test using multi-turn conversations from the LMSYS-Chat-1M dataset, with both InfiniGen and H2O configured at budget of 0.3.
\begin{figure}[th]
  \centering
  \begin{subfigure}[t]{0.489\linewidth}
    \centering
    \includegraphics[width=\linewidth]{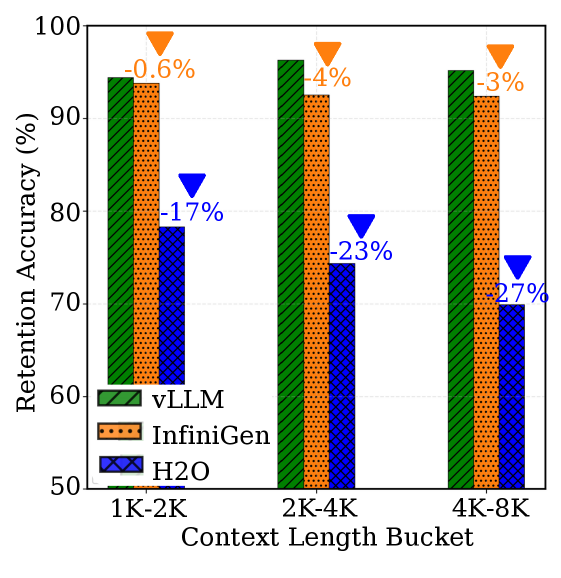}
    \caption{\centering Accuracy by context-length bucket.}
    \label{fig:retention_by_context}
  \end{subfigure}\hfill
  \begin{subfigure}[t]{0.49\linewidth}
    \centering
    \includegraphics[width=\linewidth]{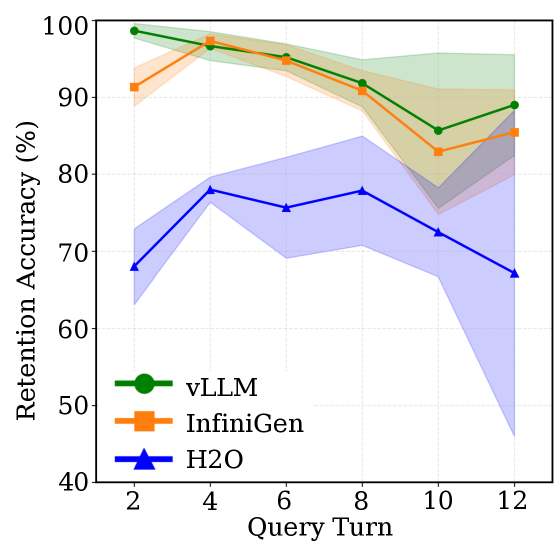}
    \caption{\centering Accuracy evolution across conversation turns.}
    \label{fig:retention_over_turns}
  \end{subfigure}
  \caption{Retention accuracy for early-context facts (Llama-3.1-8B, LMSYS-Chat-1M). (a) Accuracy by context-length bucket. (b) Accuracy evolution across conversation turns.}
  \label{fig:retention_results}
  \vspace{-1.2em}
\end{figure}

We partition conversations into three context-length buckets: 1K-2K, 2K-4K, and 4K-8K tokens. At the beginning of each conversation, we inject a verifiable fact (e.g., "My dog's name is Barkley") as a user message, followed by an assistant acknowledgment. This placement ensures the fact appears early in the context, maximizing the distance between injection and retrieval. We then process the conversation turn-by-turn, generating model responses at each exchange. At predetermined query points (turns 2, 4, 6, 8, 10, 12), we pause to ask a retention question corresponding to the injected fact. This progressive querying measures how retention degrades as context grows and the fact recedes further into the KV cache. Across 50 conversations per bucket, this yields several hundred retention tests per method, with the exact count depending on conversation lengths. We evaluate correctness by checking whether the model's response contains the expected answer, directly comparing retention across compression strategies.

Figure~\ref{fig:retention_results} presents retention accuracy across context lengths and conversation turns. As shown in Figure~\ref{fig:retention_by_context}, vLLM maintains consistently high retention (94-96\%) across all context buckets, serving as our baseline. InfiniGen closely tracks vLLM, achieving 92-94\% accuracy with only 1-4\% degradation relative to the baseline. In contrast, H2O exhibits substantial accuracy loss, retaining only 70-78\% of injected facts, a 17-27\% reduction compared to vLLM. Notably, H2O's degradation worsens with increasing context length, dropping from 17\% below the baseline at 1K-2K tokens to 27\% below at 4K-8K tokens. Figure~\ref{fig:retention_over_turns} reveals how retention evolves as conversations progress. While vLLM and InfiniGen maintain overlapping accuracy bands throughout (85-98\%), H2O consistently underperforms by 20-30 pp across all query turns. The shaded regions represent variance across context buckets, with all methods showing increased variance at later turns. These results demonstrate that InfiniGen's selective compression preserves critical early-context information, whereas H2O's eviction strategy can discard essential tokens.

\section{Discussion}
\label{sec:discussion}

Our evaluation reveals that KV cache management involves navigating a three-way trade-off between GPU memory consumption, throughput, and accuracy with no single framework dominating across all dimensions.

\textbf{When to choose vLLM.} For deployments where GPU memory is sufficient and throughput paramount, vLLM remains the clear choice. Its native FlashAttention-2 integration delivers the lowest TTFT, highest decode throughput, and flat latency scaling across batch sizes. The pre-allocated full cache eliminates management overhead entirely, making vLLM ideal for latency-sensitive applications.

\textbf{When to choose H2O.} When GPU memory is constrained but throughput requirements remain high, H2O offers an attractive trade-off. H2O achieves  70\% memory reduction while maintaining throughput comparable to vLLM, and substantially higher than InfiniGen. However, practitioners should be aware of H2O's accuracy limitations: permanent eviction disproportionately affects knowledge-intensive tasks and retention of early-context information. H2O is better suited for applications where occasional fact retrieval errors are tolerable.

\textbf{When to choose InfiniGen.} InfiniGen occupies a specialized niche: deployments requiring sparsification with minimal accuracy degradation, where latency constraints are relaxed. Its dynamic selection preserves early-context retention equally well as the full cache, a significant advantage for multi-turn conversations or applications requiring reliable fact recall. However, the throughput penalty of an order of magnitude from CPU-GPU transfers limits its practicality for high-throughput scenarios. InfiniGen becomes more attractive as interconnect technology improves; NVLink-C2C's 7$\times$ bandwidth improvement over PCIe Gen 5 could substantially narrow the performance gap.

\section{Related Work}
\label{sec:related}
We organize related KV cache management frameworks according to the three paradigms evaluated in this study.

\textbf{Memory Management Frameworks (vLLM Paradigm)}:
This paradigm optimizes memory allocation and scheduling without evicting cache entries. ORCA~\cite{yu2022orca} enables continuous batching at individual decoding steps, improving GPU utilization by allowing new requests to join ongoing batches. vTensor~\cite{xu2024vtensor} extends virtual memory abstractions to GPU tensors, enabling dynamic defragmentation without interrupting computation.

\textbf{Static Sparsification Frameworks (H2O Paradigm)}:
These frameworks reduce cache size through permanent token eviction based on attention scores. Scissorhands~\cite{NEURIPS2023_ScissorHands} retains ``pivotal'' tokens with high cumulative attention scores. StreamingLLM~\cite{xiao2023efficient} exploits the ``attention sink'' phenomenon, maintaining only initial tokens and a sliding window for theoretically infinite sequence lengths. ALISA~\cite{ISCA24_ALISA} combines dynamic token selection with optional CPU offloading, while SnapKV~\cite{li2024snapkv} performs one-time compression after prefill using an observation window.

\textbf{Dynamic Selection Frameworks (InfiniGen Paradigm)}:
This paradigm retains full KV caches (typically on CPU) and dynamically selects entries to load at each decode step. InfLLM~\cite{xiao2024infllm} uses block-level indexing across CPU-GPU hierarchy. Quest~\cite{tang2024quest} refines block selection using min-max key representations, while PQCache~\cite{zhang2025pqcache} applies Product Quantization for sub-linear retrieval. SqueezedAttention~\cite{hooper2025squeezed} employs K-means clustering to load semantically relevant key groups.

\textbf{Other Complementary Techniques}:
Orthogonal approaches include KV cache quantization, such as KVQuant ~\cite{NEURIPS2024_KVQuant}, and scheduling, such as Llumnix~\cite{OSDI24_Llumnix} for live migration. These techniques can be layered to the paradigms we evaluate.

\textbf{Characterization Efforts}: 
Since KV cache management for LLM inference is a relatively new area, only a few works have systematically explored trade-offs between frameworks. Ye et al.~\cite{ipdps25_kv_concurrency} analyzed performance bottlenecks arising from KV cache contention under concurrent workloads, finding that chunking during prefill can degrade throughput when forward passes are fast, and that eviction strategies depend on model size, memory pressure, and workload characteristics. Prior work has also analyzed individual frameworks: Kwon et al.~\cite{SOSP23_vLLM} compared vLLM against ORCA, and Li et al.~\cite{li2024survey} provided a theoretical taxonomy, but these works do not evaluate fundamentally different management paradigms.

\section{Conclusion}
\label{sec:conclusion}

We compare three representative KV cache management frameworks: vLLM (memory management), H2O (static sparsification), and InfiniGen (dynamic selection). We evaluate time-to-first-token, throughput, latency, memory utilization, and accuracy across batch sizes, context lengths, and sparsity budgets. Our findings clarify the trade-offs among these approaches under different memory, performance, and accuracy constraints. vLLM delivers the highest throughput and lowest latency when GPU memory is sufficient. H2O achieves the best memory efficiency, reducing GPU consumption by up to 70\% while maintaining competitive throughput, though at the cost of accuracy on retention-sensitive tasks. InfiniGen preserves accuracy most effectively among sparsification approaches but incurs significant throughput penalties due to costly CPU-GPU transfers in its hierarchical placement design. For reliable early-context retention, InfiniGen substantially outperforms H2O. As CPU-GPU bandwidth improves, hierarchical approaches may become increasingly viable for throughput-sensitive workloads.

\bibliographystyle{IEEEtran}
\bibliography{kvcache}

\end{document}